\documentstyle[twoside,fleqn,espcrc2]{article}

\def\eg{{\it e.g.,\ }}
\def\ie{{\it i.e.,\ }}
\def\et{{\it et al.}}

\def\vs{{\it vs.}}
\def\cO{{\cal O}}
\def\tr{{\rm tr}}

\def\gtwid{\raise.3ex\hbox{$>$\kern-.75em\lower1ex\hbox{$\sim$}}}
\def\ltwid{\raise.3ex\hbox{$<$\kern-.75em\lower1ex\hbox{$\sim$}}}
\newcommand{\ltsim}{\mathrel{\lower4pt\hbox{$\sim$}}
\hskip-12pt\raise1.6pt\hbox{$<$}\;}
\newcommand{\gtsim}{\mathrel{\lower4pt\hbox{$\sim$}}
\hskip-12pt\raise1.6pt\hbox{$>$}\;}
\def\prl#1{Phys.\ Rev.\ Lett.\ {\bf #1}}
\def\prd#1{Phys.\ Rev.\ {\bf D#1}}
\def\plb#1{Phys.\ Lett.\ {\bf #1B}}
\def\npb#1{Nucl.\ Phys.\ {\bf B#1}}

\def\seillac{Nucl.\ Phys.\ {\bf B} (Proc.\ Suppl.) {\bf 4} (1988)}
\def\fermilab{Nucl.\ Phys.\ {\bf B} (Proc.\ Suppl.) {\bf 9} (1989)}
\def\capri{Nucl.\ Phys.\ {\bf B} (Proc.\ Suppl.) {\bf 17} (1990)}

\def\tsukuba{Nucl.\ Phys.\ {\bf B} (Proc.\ Suppl.) {\bf 26} (1992)}
\def\amsterdam{Nucl.\ Phys.\ {\bf B} (Proc.\ Suppl.) {\bf 30} (1993)}

\newcommand{\fbst}{$f_B^{\rm stat}$}
\newcommand{\fb}{$f_B$}
\newcommand{\bkg}{$B \to K^* \gamma$}

\newcommand{\zast}{$Z_A^{\rm stat}$}

\newcommand{\AmS}{{\protect\the\textfont2
  A\kern-.1667em\lower.5ex\hbox{M}\kern-.125emS}}

\title{Heavy-light and light-light weak matrix elements on the lattice
\thanks{Review presented at {\it Lattice '93}, Dallas, Oct.\ 
12--16, 1993}}

\author{Claude W.\ Bernard\address{Department of Physics, CB1005, \\
        Washington University, \\ 
        St.\ Louis, MO 63130, USA}}
       
\begin{document}

\begin{abstract}
I review recent developments in lattice weak matrix element
calculations.   I focus on on $f_B$ (both with propagating quarks
and in the
static limit for the $b$ quark),
semi-leptonic form factors for $D$ meson
decay,
form factors for $B \to K^* \gamma$, and
$B_K$.
\end{abstract}

\maketitle

\section{INTRODUCTION}

The main obstacle to precise determinations of the parameters
of the Standard Model is the difficulty in calculating
hadronic matrix elements of weak operators.  The wide range in allowed
values for hadronic quantities such as 
$f_B$ (the pseudoscalar decay constant
of the $B$ meson), the form-factors for semi-leptonic
$B$ decay, and 
$B_K$ (describing mixing in
the $K$-$\bar K$ systems)
results in large uncertainties in our determinations of  the
Cabibbo-Kobayashi-Maskawa angles and $m_t$ \cite{PHENOM}.  
Lattice
gauge theory, at least in principle, should
allow us to compute such matrix elements from first
principles, \ie with control on
all sources of systematic error. 
 Here, I review the current
status of this effort.

The talk is organized as follows.  I first discuss $f_B$, both in the
static approximation \cite{EICHTEN_LAT87} and with propagating quarks. 
In the case of
\fbst, the results show a wide spread, both among various
groups working on the problem, and as a function of the lattice spacing $a$.  I explain
the reasons for the spread and show that the $a$-dependence
seems to  be understandable in perturbation theory.  The variation
among groups is likely due to the poor signal-to-noise
properties of static-light propagators.  This can introduce
spurious dependences on the details of how the signal
is extracted.  I discuss in particular the quality of the
data obtained with different
types of ``smearing.''  In the case of \fb\ with propagating quarks,
the results (at least the raw data)
from different groups seem rather better in agreement.
The main issues which I consider are the proper normalization of 
quarks with $ma\sim1$, the consistency between Wilson quarks and
Sheikholeslami-Wohlert (SW) improved quarks, and the consistency
between the propagating and static decay constants. 

Recent results
for the  form factors for semi-leptonic decays of $D$ mesons are then discussed
\cite{KENWAY}.
I focus  on the normalization
factors for the local and point-split currents and describe
how their mass dependence may be understood, at least qualitatively. 
This provides a way to reconcile (or at least reduce the discrepancies
between) results from different currents with different heavy
quark masses.

I then describe results presented at this conference
for the form factors of the rare decay $B \to K^* \gamma$.
This is a promising new way for the lattice
to provide a handle on the Standard Model.  The results suggest
that rather stringent constraints on the Standard Model could
be obtained with modest improvements in the lattice calculations
and/or the experiment.

Finally I review some exciting
new  developments in the computation of $B_K$ on the lattice.
All systematic errors but one appear to be under control,
and that one is a chiral loop effect which should be no larger
than a few percent.  The result, $B_K(\mu=2{\rm GeV})= 0.616\pm0.020\pm0.017$
or $\hat B_K=0.825\pm0.027\pm0.023$ \cite{SHARPE},  is an important contribution
of lattice QCD to Standard Model phenomenology.

\section{\fbst}

Table~\ref{tab:fbstat} gives some results for \fbst\ from various groups.  
Note that the values range from a high of $370\pm40$ MeV to a low of
$228\pm21$ MeV.  However, it is misleading to compare these
final answers, because different groups use 1) different values for
\zast, the lattice renormalization constant for the static-light
axial vector current, and 2) different methods to set the lattice scale, $a$.
Since what is really being calculated is $(f\sqrt{m})^{\rm stat}a^{3\over2}$,
the result for \fbst\ is particularly sensitive to the value used for $a$.

\begin{table}
\caption{Results for \fbst\ from various groups}
\label{tab:fbstat}
\centering
\begin{tabular}{|c|c|c|} \hline
group & $\beta$ & \fbst\  (MeV) \\
\hline
FNAL\cite{FNAL}    & 5.7 & $305(15)$ \\
                   & 5.9 & $269(20)$ \\
                   & 6.1 & $228(21)$ \\
                   & 6.3 & $237(22)$ \\
\hline
BLS\cite{BLS}  & 6.3& $235(20)\pm21$  \\
\hline
APE\cite{APE_FBSTAT,ALLTON}  & 6.0& $350(40)\pm30  $\\
                   & 6.0 [clover] & $370(40)$ \\
\hline
UKQCD\cite{UKQCD_FB}  & 6.0 [clover]& $286{}^{+8}_{-10}{}^{+67}_{-42} $\\
                      & 6.2 [clover]& $253{}^{+16}_{-15}{}^{+105}_{-14}  $\\
\hline
P-W-C\cite{PWC_FBSTAT}  & 5.74,6.0,6.26& $230(22)\pm26  $\\
                      & (extrap.)& \\
\hline
Hashimoto\cite{HASHIMOTO}  & 6.0 [nrqcd]& $320(30)\pm60$  \\
\hline
\end{tabular}
\vspace{-0.15truein}
\end{table}

\begin{figure}[tb]
\vspace{2truein}
\includegraphics{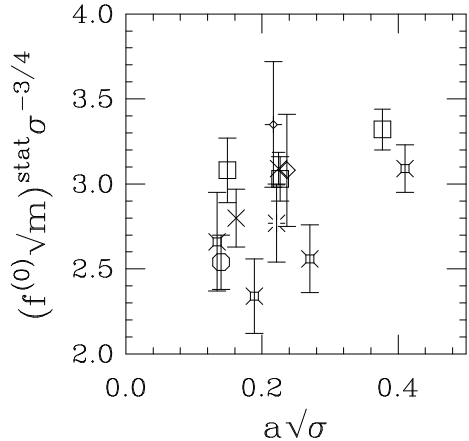}
\caption{Raw value of 
$(fm^{1\over2})^{\rm stat}\sigma^{-{3\over4}}$ \vs\ 
$a\sigma^{1\over2}$
from various groups; only the relative perturbative
correction between SW and Wilson quarks is included. 
Statistical errors only. Some points have been moved
slightly horizontally for clarity. Key:
FNAL fancy square; BLS octagon;
UKQCD (SW action) cross;
P-W-C square; APE diamond;
APE (SW action) fancy diamond;
Hashimoto burst. For references
see Table~1.}
\label{fig:fbstat-raw}
\vspace{-0.15truein}
\end{figure}

In order to make a more realistic comparison, I therefore fix
the scale uniformly by using computations \cite{BALI}
of the string tension, $\sigma$,
and plot in 
Fig.~\ref{fig:fbstat-raw} the ``raw'' value of 
$(f\sqrt{m})^{\rm stat}\sigma^{-{3\over4}}$ \vs\ $a\sqrt{\sigma}$. 
Here, ``raw'' means before multiplying by
\zast, except that in the case of the SW action, I do correct
for the difference between \zast\ with the Wilson and SW actions by using
perturbation theory \cite{ZASTAT,ZASTATSW} with
a boosted coupling \cite{LEPMAC} $g^2=1.77$ at $\beta=6.0$ and 
$g^2=1.65$ at $\beta=6.2$. 

Two features stand out in Fig.~\ref{fig:fbstat-raw}.  First, there
are real disagreements between the groups.  For the moment, I will
focus on the lower tier of results (including the FNAL and the BLS points).
I do this not only because
it includes my own work \cite{BLS}, but also because the FNAL
group uses the best sources \cite{EICHTEN_LAT91,FNAL_GOODSOURCE}.
(The UKQCD results 
are somewhat higher, but only because of the relative
perturbative correction for the SW action.)
The second
feature is the strong $a$ dependence of the lower results.
This is puzzling at first glance.  If the cause were the presence of higher
dimension operators (\cO($a$) effects), one would expect that
that the SW ($\cO(a)$ improved) points would show significantly
less scale dependence, which does not appear to be the case.
Furthermore, the results for propagating quarks seem to
show considerably smaller scaling violations. (Of course,
the size of statistical errors makes definitive statements
difficult.)  I believe, however, that most, if not all, of the
$a$ dependence can actually be explained by the ``improved
perturbation theory'' ideas of Lepage and Mackenzie.

\subsection{Tadpole improved perturbation theory and the static limit}
\label{sec:statscale}

Lepage and Mackenzie  \cite{LEPMAC} argue that bare lattice
perturbation theory is ill-behaved because 1) the bare lattice
coupling is a poor choice of an expansion parameter, and 2) lattice tadpoles
are large.  They cure the first disease by using a more physical
coupling constant such as $g_V$, which is defined in terms of the
heavy-quark potential.  The second disease is cured by
summing up the tadpoles with a 
mean field theory approach:  they show that the simple replacement of
the link $U$ by its mean value $u_0$ accounts for the bulk of
the perturbative corrections to many quantities.  For example,
for Wilson fermions, the replacement $U\to u_0$ just changes
the hopping parameter $\kappa$ to $\tilde \kappa \equiv \kappa u_0$.
The resulting free field theory has a critical hopping parameter
$\tilde \kappa_c = {1\over 8}$ which would imply
$\kappa_c = {1\over 8 u_0}$, a relation which agrees reasonably
well with simulations when one defines
$u_0 \equiv \langle {1\over 3} \tr U_{\rm plaq}\rangle^{1\over4}$.
In the following, I will define $u_0$ by $\kappa_c$ for Wilson
quarks; that is,
I take
\begin{equation}
\label{eq:unaught}
u_0\equiv {1\over 8\kappa_c} = 1 -0.109 g^2 + \dots\ .
\end{equation}

In computing some perturbative quantity, the prescription for
``mean-field improvement'' (or ``tadpole improvement") 
is then to take out explicit powers of $u_0$ (depending on the
number of links which enter the definition of the object),
replace them by the value of $u_0$ in the simulation ($1/(8\kappa_c)$),
and remove the appropriate power of $1 -0.109 g^2 + \dots$ from
the remaining perturbative expansion.  For example,
the 1-loop renormalization factor for the local axial current with
Wilson fermions in the chiral limit is \cite{MART-ZHANG}
$\tilde Z_A = 2\kappa_c(1 -0.133g^2)$,  where the $\tilde{}$ on 
$\tilde Z_A$ implies that I include the standard factor $2\kappa_c$ relating
Wilson to continuum fermions.  The tadpole-improved prescription
would then be
\begin{equation}
\label{eq:zahat}
\tilde Z_A \to 2u_0\kappa_c{1-.133g^2\over 1-.109g^2}= {1\over4}(1-.024g^2_V({1\over a}))
\end{equation}
The factor of $u_0$ which is removed is really two factors of $\sqrt{u_0}$
coming from wave function renormalization on the external lines.
The small coefficient in front of $g^2_V$ indicates that improved perturbation
theory is working well \cite{COMMENT}.  The scale ($1/a$) 
at which $g^2_V$ is evaluated
is a typical one for non-tadpole diagrams;
when the quadratically divergent tadpoles are left in, the scale
is typically closer to $\pi/a$.

An interesting feature of perturbative quantities in the static
approximation, however, is that they are much less tadpole-dominated
than the corresponding quantities with Wilson quarks. 
This can easily be understood by examining the static action:
\begin{eqnarray}
\label{eq:Sstat}
S_s= \sum_x  \bar h_x ({1 + \gamma_0\over 2})[h_x - 
U_0^\dagger(x-\hat0)
h_{x-\hat0}] \\
\to \sum_x  \bar h_x ({1 + \gamma_0\over 2})[h_x - u_0h_{x-\hat0}] 
\end{eqnarray}
From eq.~(4) one can easily calculate, using a hopping expansion,
 the mean field theory static propagator. We get
\begin{equation}
\label{eq:Gstat}
G_s(x,y) = \delta^{(3)}_{\vec x,\vec y}\  e^{-(\ln u_0^{-1})(t_x-t_y))}
\end{equation}
This shows that $u_0$ contributes to mass renormalization ($\ln u_0^{-1}$),
but not wave function renormalization, since there is no overall factor in
$G_s$.

Equivalently, in momentum space one has $G_s(p)= (1-u_0e^{ip_0})^{-1}$.
The $u_0$ factor looks like wave function renormalization at first
glance. However, expanding about the pole, we let $p_0=i\ln u_0 + p'_0$
and find $G_s(p'_0) = (1-e^{ip'_0})^{-1}$, which still has residue 1.
Thus,
unlike the Wilson case,
there is  no tadpole contribution to the
wave function renormalization of heavy quarks;
one can also see this explicitly in the perturbative calculations
\cite{ZASTAT}. 

In the case of the static-light axial current renormalization constant,
one then takes out only a factor of $\sqrt{u_0}$ for the
light quark, and divides by $1- {.109\over2}g^2$. Using the
known perturbative results \cite{ZASTAT,ZASTATSW}, I then get, 
in the chiral limit for the light quark, 
\begin{equation}
\label{eq:zastattad}
\tilde Z^W_{A,{\rm stat}}
\to ({1\over2})\left[1-g_V^2\left(q^*\right)\big(.135-\delta\big)\right]
\end{equation}
\begin{equation}
\label{eq:zastattadsw}
\tilde Z^{SW}_{A,{\rm stat}} \to 
\sqrt{\kappa_c^{SW}\over 4 \kappa_c^W}
\left[1-g_V^2\left(q^*\right)\big(.090-\delta\big)\right]
\end{equation}
where the superscripts ``$W$'' and ``$SW$'' on $\tilde Z_{A,{\rm stat}}$
indicate Wilson or SW light quarks, respectively, 
$\delta\equiv .025\ln\left(am_B\right)$, and $q^*$ is the 
scale at which $g^2_V$ should be evaluated.
The static renormalization constants
are clearly not tadpole dominated,
since the coefficients of $g^2_V$ in the above equations are
relatively large (compare eq.~(\ref{eq:zahat})).

It is not completely clear what scale $q^*$ to put into
eqs.~(\ref{eq:zastattad}),(\ref{eq:zastattadsw}). In cases where the scale
after removing tadpoles has been estimated
using the method of ref.~\cite{LEPMAC}, it is $\ltwid 1/a$:
for $\kappa_c$, one has $q^*=1.03/a$ \cite{LEPMAC}; for
two non-relativistic QCD (NRQCD) masses, 
one has $q^*=0.81/a$ and $ 0.67/a$ \cite{MORNINGSTAR}.
In the current case, the scale is estimated
to be $2.36/a$ before tadpole removal
\cite{HILL}, but it may not be reduced as much
as in the other cases by removing the tadpoles, since the tadpoles play a less
important role.  
Indeed, since the time when this talk was given, the
non-tadpole
scale has been estimated by Hill 
and Hernandez \cite{HILL_INPREP}; their preliminary result
is $q^*= 2.18/a$. 
Note that
$q_V^2(2.18/a)$ is quite large, ranging from 2.8 at $\beta=5.7$ to
1.9 at $\beta=6.3$. 
This makes the term $g_V^2\left(q^*\right)\left(.135-\delta\right)$ range from
$.26$ to $.23$; with $q^*=1/a$, the corresponding numbers are $.42$ to $.31$.
The higher order perturbative effects could therefore
be quite large, perhaps 10\% or even more, which implies
considerable uncertainty in \fbst.

\begin{figure}[tb]
\vspace{2truein}
\includegraphics{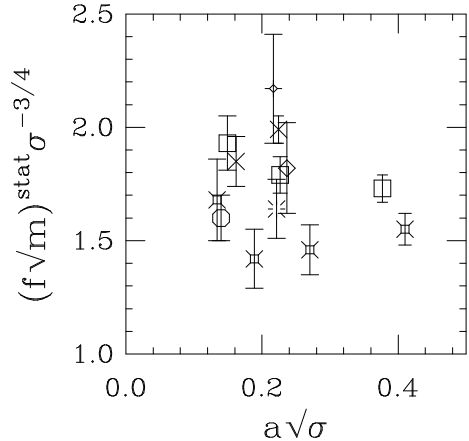}
\caption{Same as Fig.~1 but
including the tadpole improved perturbative
corrections, with $g_V^2$ evaluated
at  scale $q^*=1/a$.}
\label{fig:fbstat-ainv}
\vspace{-0.15truein}
\end{figure}

\begin{figure}[tb]
\vspace{2truein}
\includegraphics{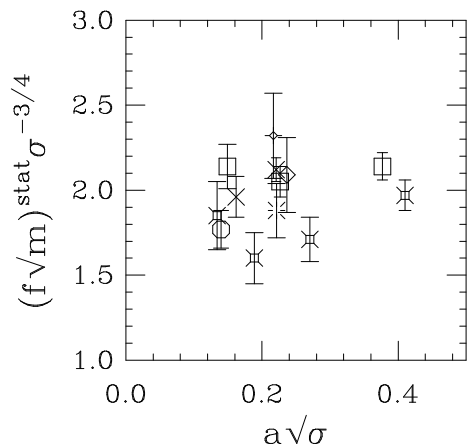}
\caption{Same as Fig.~2 but
at  scale $q^*=2.18/a$.}
\label{fig:fbstat-twoainv}
\vspace{-0.15truein}
\end{figure}

In Figs.~\ref{fig:fbstat-ainv} and  \ref{fig:fbstat-twoainv}, I put
in the tadpole improved corrections to the raw static numbers,
using $q^*=1/a$ and $q^*=2.18/a$ respectively.
Note that with $q^*=1/a$, the $a$ dependence of the the lower tier
of results is small. With $q^*=2.18/a$, there is still may be significant $a$
dependence, but it is reduced from Fig.~\ref{fig:fbstat-raw}, and the
results are consistent with a constant if one drops the point at
largest $a$  ($\beta=5.7$).
One is now tempted to extrapolate to $a=0$.  With either scale,
a linear extrapolation of the FNAL and BLS results
gives $(f\sqrt{m})^{\rm stat}\sigma^{-{3\over4}} \approx 1.6\pm0.2$
at $a=0$, which would imply $f_B^{\rm stat}\approx 190\pm25$ MeV.
Extrapolation of the UKQCD (SW-action) points would produce
almost the same answer.
Clearly, however, there are large systematic errors 
in the result
at this stage.  I estimate that total systematic
error associated with the extrapolation, with the setting of the scale,
with the uncertainty in the perturbative corrections, and with
the extraction of the raw numbers themselves (see below) is
roughly twice the statistical error.

\subsection{Smearing}
I now come to the issue of the large variation in the
results from different groups. I believe this
stems from the intrinsically poor signal-to-noise ratio for static-light
propagators at large times \cite{LEPAGE_LAT91,BLS_LAT91,EICHTEN_LAT91}. 
The problem has been dealt with by smearing the static quark 
(or both quarks) in
the static-light sources. This can improve the situation in two ways:
1) by producing a better overlap with the ground state, thereby moving the
plateau in to smaller times where the signal-to-noise is better,
and 2) by increasing the statistics.  The second point, which has not
always been appreciated in the literature, will be discussed below.  

\fbst is computed from the asymptotic behavior of the correlation
functions
\begin{eqnarray}
\label{eq:LS}
G_{LS}(t)= \sum_{\vec x} \langle0|A_0(\vec x,t) 
                \chi^\dagger_S(\vec0,0)|0\rangle \\
\label{eq:LSeqSL}
        {\rm ``}={\rm "}  \sum_{\vec x} \langle0| \chi_S(\vec x,t) A_0^\dagger
        (\vec 0,0)|0\rangle \ ,  \\
G_{SS}(t) =  \sum_{\vec x} \langle0|\chi_S(\vec x,t) 
                \chi^\dagger_S(\vec0,0)|0\rangle \ ,
\end{eqnarray}
where $A_0 = \bar h \gamma_\mu \gamma_5 q$ is the local or almost
local (\ie point split) axial current ($h$ and $q$ are the
static and light quark fields, respectively), and $\chi_S$ is the smeared
meson interpolating operator:
\begin{equation}
\chi_S(\vec x,t)= \sum_{\vec y} f(\vec y) \bar h(\vec x+\vec y,t) \gamma_5 
                   q(\vec x,t) \ ,
\end{equation}
with $f(\vec y)$ some smearing function. (I assume here that only the static quark is
smeared, which is the most common practice.)

I write ${\rm ``}={\rm "}$ in eq.~(\ref{eq:LSeqSL}) because the two
correlators, while equal when averaged over an infinite number of
configurations, have different statistical errors with a finite
number of configurations. 
The smeared-source, local-sink, eq.(\ref{eq:LS}), is considerably
less noisy than the local-source, smeared-sink, eq.(\ref{eq:LSeqSL}).
This was noticed with the first use of
static-light smearing \cite{EICHTEN_LAT89}; a nice comparison
of the results from the two appears in the recent UKQCD paper \cite{UKQCD_FB}.

The reason for the difference in statistics is easy to understand.
Since the static quark propagates along a straight line in the time
direction, 
only the static propagator from $(\vec 0,0)$ to $(\vec 0,t)$ 
(coming only from terms with $\vec x= -\vec y$ in eq.~(\ref{eq:LSeqSL}))
contributes when the source is local.
Independent of what the smearing function is, the result
just depends on a single product of links in each configuration,
and hence has large fluctuations.  When the source is smeared,
many products of links enter into the result from each configuration,
and the fluctuations are thereby greatly reduced. 

Note that the statistics issue is completely independent of the
the relative ground-state \vs\ excited-stated overlap for
a given interpolating field.
This means there may be a certain amount of tradeoff between good
overlap and good statistics.  For example, wall sources, in which
both the static quark and the light quark are smeared over the entire lattice,
clearly have poor overlap, since the mean distance between the static
and light quarks is much larger than the size of the ground state.
However, they have potentially very good statistics since
all time-like link products on the lattice are used.

Figs.~\ref{fig:apeSS} and \ref{fig:apeLS-SS} show data 
for the  $G_{SS}$ effective mass and for $G_{LS}/G_{SS}$ from 
APE \cite{APE_FBSTAT} with the SW action at $\beta=6.0$. 
The heavy quark is
smeared over a cube of side $L_S$; various values of $L_S$
are shown.  $G_{LS}$ is calculated with local-source, smeared-sink
(eq.~(\ref{eq:LSeqSL})), which accounts the comparatively
large errors in Fig.~\ref{fig:apeLS-SS} (compared with, say, the
UKQCD data \cite{UKQCD_FB}), despite the
large number of configurations (210). I find it difficult to see
consistent, convincing plateaus in both $G_{SS}$ and $G_{LS}/G_{SS}$:
$L_S=7$ and $L_S=9$ are reasonable but $L_S=5$ is 
somewhat doubtful in the former; 
while only $L_S=5$ is convincing in the latter. 
One really needs plateaus
for the same value of $L_S$ in both correlators to be able to extract
an answer.
The APE group takes
the variance between the $L_S=5$ and the $L_S=7$ results as an estimate
of the systematic error here; however I am not convinced that this
is a good estimator of the error.  

\begin{figure}[tb]
\vspace{2truein}
\caption{APE data for the 
effective mass from the smeared-smeared correlator for various
smearing sizes, $L_S$.  $\beta=6.0$, SW action, 210 configs.}
\label{fig:apeSS}
\vspace{-0.15truein}
\end{figure}

\begin{figure}[tb]
\vspace{2truein}
\caption{APE data for the 
ratio $G_{LS}/G_{SS}$ for various smearing sizes, $L_S$.  
$\beta=6.0$, SW action, 210 configs.}
\label{fig:apeLS-SS}
\vspace{-0.15truein}
\end{figure}

Fig.~\ref{fig:wup} shows the effective mass for various
smearings in  $G_{SS}$ from
the PSI-Wuppertal-CERN group (P-W-C) \cite {PWC_FBSTAT}.
I do not find the plateau, which they consider to start at $t=3$,
very convincing.  Since their analysis determines the mass from this
correlator and then uses it in determining the location of the
plateau in $G_{LS}$, I suspect there may be considerable  systematic error
involved here.  I note, however, that the P-W-C group  has  varied
the parameters of their analysis and does not find large variations in
the results. A possible additional source of systematic error
arises from the rather small size of the lattice here ($18^3 \times 48$
at $\beta=6.26$).

\begin{figure}[tb]
\vspace{2truein}
\caption{P-W-C data for the effective mass from $G_{SS}$, for various
smearings. $18^3 \times 48$, $\beta=6.26$, $\kappa_{\rm light} = .1492$,
43 configs.}
\label{fig:wup}
\vspace{-0.15truein}
\end{figure}

\begin{figure}[tb]
\vspace{2truein}
\includegraphics{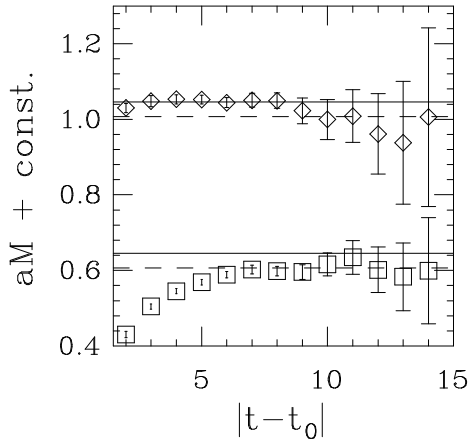}
\caption{BLS data for the effective mass from $G_{SS}$ (diamonds)
and $G_{LS}$ squares.
$24^3 \times 39$, $\beta=6.0$, $\kappa_{\rm light} = .155$,
8 configs. The diamonds have been displaced by $+0.4$ for clarity.
The dashed lines and solid lines are different simultaneous
fits to the correlators; see text.}
\label{fig:bls-six}
\vspace{-0.15truein}
\end{figure}

Some of the difficulty in extracting \fbst\ from simulations is illustrated
in fig.~\ref{fig:bls-six} from our group \cite{BLS}.
The data, at $\beta=6.0$ with cube smearings of side $L_S=9$, is fit
in two different ways:
The dashed lines show  simultaneous fits to
$G_{SS}$ and $G_{LS}$ in the range $t=(5,10)$; while the solid
lines show  simultaneous fits to
$G_{SS}$ in the range $(3,7)$ and
$G_{LS}$ in the range $(9,13)$. 
While both sets of fits appear,
at first glance, reasonably convincing, they give vastly
different results. Indeed \fbst\ extracted from the first (dashed-line) fits
is $\approx 1.5$ times larger than that extracted from the second (solid-line) fits.
For this reason we have not felt confident of quoting a result
for \fbst\ at $\beta=6.0$.
Note that, although the number of configurations is small (8),
the statistical errors in the data with  our sources
are at least as small as those
in the data of the previously discussed groups.

\begin{figure}[tb]
\vspace{2truein}
\includegraphics{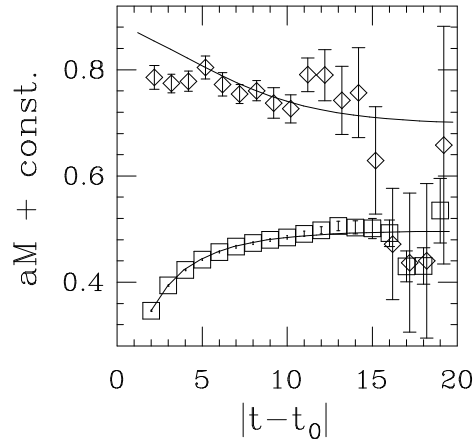}
\caption{BLS effective masses 
with wall sources at $\beta=6.3$, $24^3 \times 55$, 20 configs.\ (40 sources),
$\kappa_{\rm light} = .150$.  Diamonds: $G_{SS}$ (displaced $+0.2$ for
clarity); squares: $G_{LS}$.  A simultaneous two-exponential fit to both channels
is used. The range is $(5,13)$ for $G_{SS}$; $(3,12)$ for $G_{LS}$.}
\label{fig:bls-ptwall}
\vspace{-0.15truein}
\end{figure}

Fig.~\ref{fig:bls-ptwall} shows our data \cite{BLS} with wall sources
at $\beta=6.3$.  Note that the errors in $G_{LS}$ are very small,
but that a two-exponential fit is required because the 
relative ground-state overlap is poor.  
For $G_{SS}$, the errors are much larger, because smearing at the
sink does not buy anything in terms of statistics, but just adds noise
coming from large separations between the quarks.
These features are all consistent with the discussion above.

\begin{figure}[tb]
\vspace{2truein}
\includegraphics{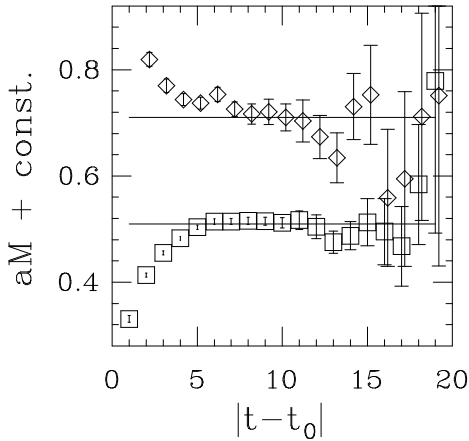}
\caption{Same as Fig.~8
but the sources are cubes ($L_S=13$) and a simultaneous
single exponential fit in the range $(9,16)$ is used.}
\label{fig:bls-ptcube}
\vspace{-0.15truein}
\end{figure}

Our data with cube sources at $\beta=6.3$ are
shown in Fig.~\ref{fig:bls-ptcube}. The errors in 
$G_{LS}$ are larger than with the wall sources, but the
ground state overlap is better, and a good plateau is seen.
The plateau for $G_{LL}$ is less convincing, but it does
have a mass
which is consistent with that of the other channel (the the channels
are fit simultaneously).
We feel safer with the cube sources than with 
the walls because we can use single-exponential fits,
 and we extract our final results from cubes of side $L_S=15$.
Note, however, that the wall sources, as well as a range of 
different cube sizes, give consistent results. This data
set is also stable under change in fitting intervals;
unlike the $\beta=6.0$ case, such shifts change the results by only
$\approx 3\%$.

Data from the FNAL group \cite{FNAL} is
shown in Fig.~\ref{fig:fnal}.  The sources
used here are close to ideal: their multistate fitting method
\cite{EICHTEN_LAT91,FNAL_GOODSOURCE}
gives a source which is nearly the ground state
wavefunction.  Despite this, however, it appears  that the results
could be changed appreciably
by moving the fitting intervals.
Admittedly, this is
not a ``typical'' but more like a ``worst case'' 
example. However,  it should serve to emphasize that important
systematic errors associated with extracting the ground state
exist in the data of all the groups.  It will probably take an order 
of magnitude better statistics (and good sources) 
to be really confident that this
systematic is controlled.

\begin{figure}[tb]
\vspace{2truein}
\caption{Effective masses for $G_{LS}$ (above) and $G_{SS}$ from the
FNAL group. $\beta=5.9$, $16^3\times 32$, $\kappa_{\rm light}=.157$,
48 configs.}
\label{fig:fnal}
\vspace{-0.15truein}
\end{figure}

Two new developments discussed at this conference show
some promise.  First, Draper and McNeile \cite{DRAPER}
have devised a method of constructing static-light
meson sources which is complementary to the FNAL
approach but produces similarly near-ideal sources.
They show some preliminary data with very nice plateaus and
are also able to extract cleanly the wave functions for excited
as well as ground states.  It will be interesting to see how well they can
determine \fbst.  Second, two groups \cite{DAVIES_HL,HASHIMOTO}
are using NRQCD for the heavy quark in the heavy-light system.
They seek to take advantage of the improved signal-to-noise
ratio \cite{LEPAGE_LAT91} for NR-light, as opposed to static-light,
mesons.  The results are quite encouraging: their effective mass plots
are clearly better than static-light ones with comparable sources.
Hashimoto evaluates
of \fbst\ (see Table~\ref{tab:fbstat}); Davies \et\ extract $f_B$ itself
(see below).

The NRQCD group \cite{DAVIES} has also carried through a very nice
calculation of the $b$-quark mass.
They get $m_b=4.7\pm0.1$ GeV.  This has an
important implication for the static theory.
It has been suggested \cite {MAIANI} that the $1/m_Q$ 
($Q$ is a generic heavy quark) corrections
to the static limit would not be computable due to the
presence of power-law divergences, which could induce
large non-perturbative effects.  The success of the NRQCD
calculation of $m_b$, which involves perturbative calculations
of such power-law divergences and has several cross-checks,
indicates that the concerns expressed in ref.~\cite{MAIANI} are not
likely to be a problem in practice.

I have one final remark on \fbst.
Given the issues relating to smearing and $a$ dependence
which were discussed above,
I was dismayed to find the
 following comment made in ref.\cite{APE_FBSTAT}:

{\small ``The result of [BLS --- ref.~\cite{BLS}] is much lower than all other results
obtained at a fixed value of $a$. A possible 
explanation may be found in the fact that in [ref.~\cite{BLS}] 
the results were obtained using very large 
smearings (or the wall source), which the authors
believe to be the most coupled to the lightest 
state, in contrast with the findings of the
present work and of [the FNAL group --- ref.~\cite{FNAL_LAT92}].''}

We have clearly stated, at least three 
times \cite{BLS_LAT91,BLS_LAT92,BLS}, the 
trade-off between ground-state overlap
and statistics which motivate our use of  wall sources
or ``very large'' smearings (which appear so large only when
compared to the small smearings necessitated by the use
of eq.~(\ref{eq:LSeqSL}) in \cite{APE_FBSTAT}).

\section{$f_B$ WITH PROPAGATING QUARKS}

Since to approach the $b$ quark on currently available lattices
one needs quark masses with $ma\sim 1$,  it is necessary
to study how lattice quarks behave when the graininess  of the
lattice is comparable to their Compton wavelength. 
Kronfeld and Mackenzie \cite{KROMAC,KRONFELD_LAT93} 
have initiated such a program.
For zero momentum quarks the idea is quite simple.
In a free theory the $\vec p=0$ Wilson
propagator is easily shown to be:
\begin{eqnarray}
G_{\rm latt}(t) =  {1\over 2 \kappa e^{aM_1}} \left({1+\gamma_0 \over 2}\right)
e^{-aM_1t} \\
 = {1\over 2 \kappa e^{aM_1}} G_{\rm cont}(t) \ ,
\label{eq:free-quark}
\end{eqnarray}
where the ``pole mass'' $M_1$ is related to the Lagrangian mass $m_0$
by $aM_1 = \ln(1 + am_0)$, with $am_0 = 1/(2\kappa) - 4$ for $r=1$,
and $G_{\rm cont}$ is the continuum $\vec p=0$ propagator.
The factor $2 \kappa e^{aM_1}$ can also be written as $1-6\kappa$,
and it is therefore clear that the correct normalization for
a massive free quark field is $\sqrt{1-6\kappa}$, rather than
the $\sqrt{2\kappa}$, which is only valid near the massless
limit.  

Interactions can easily be taken into account in the limit
of tadpole dominance.  In that case, one just replaces $\kappa$
by $\tilde \kappa \equiv \kappa u_0 = \kappa/(8\kappa_c)$
(see discussion around eq.~(\ref{eq:unaught})). 
In the tadpole approximation, one thus has a normalization
factor of $\sqrt{1-6\tilde\kappa}$ for each Wilson quark field
in the limit $\kappa\to 0$ ($aM_1 \to \infty$).

What are the corrections? First of all there are ones of $\cO(p^2)$.
These are small if $M_1 \gg |\vec p|$, which should be the case
for heavy-light mesons (for which $|\vec p| \sim \Lambda_{\rm QCD}$ )
on current lattices (for which
$aM_1 \sim 1$ implies $M_1 \gg \Lambda_{\rm QCD}$).
A systematic improvement procedure, similar to that of NRQCD,
can then take these higher order corrections in to account.
Second, there are $\cO(g^2)$ corrections. These are small
if tadpole dominance is a good approximation.  Some early
perturbative results seem to bear this out \cite{KRONFELD_LAT93}.
However, recall that tadpole dominance does not work well
in the static limit, so one expects some fairly large
perturbative corrections ($\sim 20\%$ to $30\%$) to show up
somewhere.

The SW action  is a systematic improvement on the Wilson action, eliminating
terms of $\cO(a)$ (but not $\cO(g^2a)$).  One therefore expects it
to include the $\cO(aM_1)$ difference between the
massless normalization ($\sqrt{2\kappa}$) and the large mass
normalization ($\sqrt{2\kappa e^{aM_1}}$).  This is indeed the case.
In the usual SW implementation, one ``rotates'' the quark field
by
\begin{equation}
\psi \to \left( 1 + {a \not\!\! D \over 2}\right) \psi = 
\left( 1 + {a m_0 \over 2}\right) +\cO(a^2) \ . 
\end{equation}
One therefore automatically includes a factor which
equals $\sqrt{e^{aM_1}}$ up to
$ \cO((am_0)^2)$. However, my suggestion to those using the SW action
would be to use the exact ($g\to0$, $\kappa\to 0$) normalization instead.
This is ``equivalent'' in that its difference
with what they are
doing already is higher order ($\cO(a^2)$), but it guarantees that the static limit
will be recovered in the $M_1 \to \infty$ limit (up to $\cO(g^2)$).

Two groups \cite{ALLTON,UKQCD_FB} have presented evidence that
the use of the $\sqrt{1-6 \tilde \kappa}$ norm brings the
Wilson fermion results into closer agreement with the
SW results, as expected.  For example, Fig.~\ref{fig:sw-wilson} \cite{UKQCD_FB}
shows $f_P\sqrt{M_P}$ ($P$ is a generic heavy-light pseudoscalar)
for
Wilson quarks with the  $\sqrt{1-6\tilde\kappa}$ normalization,
and with the (``naive'') $\sqrt{2\kappa}$ normalization, in comparison with the
SW results. The naive results vanish exponentially as
$M_P \to \infty$, as expected from eq.~(\ref{eq:free-quark}), and
therefore cannot be consistent with the static theory.  The
$\sqrt{1-6\tilde\kappa}$ results are in good agreement
with the SW results over the whole range of meson masses examined; this is
actually somewhat surprising, since the two are 
only supposed to agree up to terms of $\cO((am_0)^2)$. On the other hand,
the P-W-C group \cite{GUESKEN} has presented results showing that $f_P$
is less dependent on the lattice spacing if the naive,
rather than the $\sqrt{1-6\tilde\kappa}$,  normalization is used,
although both extrapolate to the same value as $a\to0$.   I do 
not understand the reason for these results.  It is possible that
the naive normalization may accidentally scale better in some
limited range of $M_P$, but it is clear, as Fig.~\ref{fig:sw-wilson}
illustrates,  that  the naive normalization
is completely incorrect  as $M_P\to\infty$, and 
must not scale with $a$ in that limit.

\begin{figure}[tb]
\vspace{2truein}
\caption{Comparison, from the UKQCD group, of 
results for $f_pM_P^{1\over2}$ \vs\ $1/M$.
Key:
SW action (circle);
Wilson action with $(2\kappa)^{1\over2}$ normalization
(diamond); Wilson action  with $(1-6\tilde\kappa)^{1\over2}$ normalization
(square).}
\label{fig:sw-wilson}
\vspace{-0.15truein}
\end{figure}

\begin{figure}[tb]
\vspace{2truein}
\includegraphics{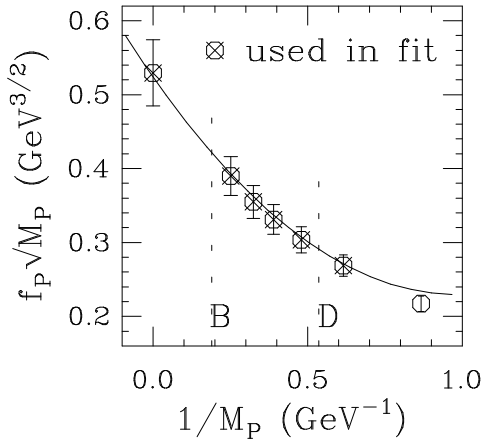}
\caption{The combined (propagating and static) analysis 
for $f_PM_P^{1\over2}/(1+{\alpha_s\over\pi}\ln(aM_P))$ at $\beta=6.3$
from the BLS
group. The propagating quarks are normalized
with $(1-6\tilde\kappa)^{1\over2}$.
The solid line is a covariant fit, quadratic in
$1/M_P$, to the marked points, and has $\chi^2/{\rm dof} = 1.9/3$. 
If the
renormalization were done exactly as advocated in sec.~2.1,
the static point would move down $3\%$  (using $q^* =2.18/a$);
the propagating point, $1.5\%$ (using $q^*=1/a$).}
\label{fig:bls-propstat}
\vspace{-0.15truein}
\end{figure}

With the  $\sqrt{1-6\tilde\kappa}$ normalization, the results
with propagating Wilson quarks seem to be in good agreement with
the static result. Fig.~\ref{fig:bls-propstat} shows the comparison
between the two methods from our calculations at $\beta=6.3$ \cite{BLS}.
The fit to both results has an excellent $\chi^2$.
Fig.~\ref{fig:ukqcd-propstat} is a similar comparison from UKQCD
\cite{UKQCD_FB}
at $\beta=6.0$
with SW propagating quarks.
Here the solid line fit to the open circles and the static point
looks  less good, but, because of the correlations between the 
propagating quark points, it is actually not 
too bad: $\chi^2/{\rm dof} = 1.5 $.
The fit would improve if tadpole-improved perturbation theory were
used for the renormalization factors $ Z_A$ of both the static and 
propagating axial currents.  For SW propagating quarks, $ Z_A$
actually is larger after tadpole improvement; while in the static
case it is reduced.  In Fig.~\ref{fig:ukqcd-propstat}, I have
attempted to show how the points would move after tadpole
improvement.  There is some uncertainty here since it is not 
completely clear how to apply tadpole improvement in the SW case 
(\eg\ whether to use the Wilson or SW $\kappa_c$ --- there is
considerable difference).  However it is clear that the agreement
between static and propagating will get better.  Use of tadpole
improvement should also reduce the large uncertainty in scale in the UKQCD 
results, which is due to a low value of $f_\pi$. Tadpole improvement
will bring up the corresponding renormalization constant, making
it closer to the non-perturbative value of $\sim1.09$ \cite{NONPERT}.
Note also that a determination of $f_B$ using non-relativistic heavy quarks
(on the same UKQCD configurations) gives a result
consistent with the other two methods \cite{DAVIES_HL}.

\begin{figure}[tb]
\vspace{2truein}
\caption{The combined analysis for 
$f_PM_P^{1\over2}/(1+{\alpha_s\over\pi}\ln(aM_P))$ at $\beta=6.0$
from UKQCD.  SW propagating quarks are used.  The dashed line is a fit
to the open circles and the open square is its intercept
at $1/M_P=0$.  The solid line is the fit with the static
point (cross) included.  The solid circles show my estimate
for the central values if tadpole improvement were used.}
\label{fig:ukqcd-propstat}
\vspace{-0.15truein}
\end{figure}

In early work \cite{ABADA}, fits of static and propagating
results seemed to show agreement between the two approaches, even
though the naive $\sqrt{2\kappa}$ normalization was used in the
propagating case.  However, such fits were non-covariant;
they did not take into 
account the correlations in the propagating-quark data.
In ref.~\cite{BLS}, we show that a reasonable-looking non-covariant
fit can be made between the static point and moderately heavy, naively
normalized, a propagating results.  However, once correlations
are included, the fit looks very bad and has $\chi^2/{\rm dof}=16/2$.
Clearly: (a) results with naively normalized propagating quarks 
are inconsistent with the static approach, and (b) covariant fits
with  meaningful $\chi^2$ are  crucial when testing the consistency
of various approaches.

I also call attention to the recent determinations
of heavy-light decays constants in full (unquenched)
QCD \cite{HEMCGC}.  The results
are generally consistent with the quenched results,
although at heavy mass (close to the $B$) there
tends to be some disagreement (the full QCD results are
higher).  This is not likely to
be the effect of quenching, especially since the
dynamical quark masses used are not very light.  Rather, current computer
limitations force the full QCD calculations to be done
at rather strong coupling.  Near the $B$, $ma$ can then
be quite large,  and the $\cO(p^2)$ corrections mentioned
above can become quite important.  In particular, the
coefficient of $p^2$ in the kinetic energy is no longer
$1/(2M_1)$, where $M_1$ is the ``pole mass,'' but
$1/(2M_2)$ \cite{KROMAC}.  Since $M_2 \gg M_1$ for
large $M_1a$, the full QCD points actually correspond
to considerably heavier meson mass. If this effect
were put in,  the disagreement between
quenched and full results would likely go away.
Note that the correction $M_1 \to M_2$ has
already been applied to the data in ref.~\cite{BLS}.

Table~\ref{tab:fbfd} shows results from various groups for
$f_B$ and $f_D$, and Table~\ref{tab:bls-ukqcd} compares
results from BLS and UKQCD for
other decay constants and
ratios.  The results are generally in quite good agreement.
At the current level of errors, differences in procedure 
do not show up significantly here. 
For example, in the case of the BLS \vs\ P-W-C
results, different normalizations
of the propagating quarks ($\sqrt{1-6\tilde\kappa}$ \vs\
$\sqrt{2\kappa}$) tend to cancel the effects of different
\fbst\ (see sec.~2.2),  producing quite similar final results for $f_B$.
The only significant difference between the BLS and UKQCD
results occurs in the ratios, especially $f_{B_s}/f_B$.  This
difference can be traced mainly to the fact that UKQCD do not include
the static values for the ratios in their analysis: both groups
get relatively low values ($1.11$, $1.14$) in the static limit.  
Because of the problems with the static signal, it is not
clear at this point which procedure is more reliable.

A result not shown in the tables is the HEMCGC value using
full QCD ($\beta=5.3$, Wilson quarks):
$f_{D_s}=345(5)\pm48$ MeV.  I do not understand the
reason for this large value compared to the quenched results.
HEMCGC suggest that it may be due to the larger lattice spacing in these
full QCD calculations.

\begin{table}
\caption{Results for $f_B$ and $f_D$ from various groups}
\label{tab:fbfd}
\centering
\begin{tabular}{|c|c|c|} \hline
group & $f_B$ (MeV) & $f_D$ (MeV) \\
\hline
\hline
BLS\cite{BLS}  & $187(10)\pm37$& $208(9)\pm37$  \\
\hline
UKQCD\cite{UKQCD_FB}  & $160{}^{+6}_{-6}{}^{+53}_{-19} $& $185{}^{+4}_{-3}{}^{+42}_{-7} $\\
\hline
P-W-C\cite{GUESKEN}  & 204(50)& $212(40)$\\
\hline
APE\cite{ALLTON}  & & $230(30)$  \\
\hline
NQRCD\cite{DAVIES_HL}  &160(10) &  \\
\hline
HEMCGC\cite{HEMCGC}  & $200(10)\pm48$ & $250(5)\pm45$  \\
\hline
\end{tabular}
\vspace{-0.15truein}
\end{table}

\begin{table}
\caption{Comparison of other BLS and UKQCD results.
Decay constants are in MeV.} 
\label{tab:bls-ukqcd}
\centering
\begin{tabular}{|c|c|c|} \hline
& BLS\cite{BLS} & UKQCD\cite{UKQCD_FB} \\
\hline
\hline
$f_{B_s}$ & $207(9)\pm40$& $194{}^{+6}_{-5}{}^{+62}_{-9} $  \\
\hline
$f_{D_s} $& $230(7)\pm35$& $212{}^{+4}_{-4}{}^{+46}_{-7} $  \\
\hline
$f_{B_s}/f_B$ & $1.11(2)\pm.05$& $1.22{}^{+.04}_{-.03}$  \\
\hline
$f_{D_s}/f_D$ & $1.11(2)\pm.05$& $1.18{}^{+.02}_{-.02}$  \\
\hline
\end{tabular}
\vspace{-0.15truein}
\end{table}

\section{SEMILEPTONIC FORM FACTORS}

Four groups have recently obtained new results for $D\to K$ and/or $D\to K^*$
\cite{ELC_SEMINEW,STELLA,RICHARDS,GUPTA}.
On the whole, the results are in agreement with previous work
\cite{ELC_SEMIOLD,BES}.  Nice plots of the 
form factors as functions of $q^2$ (the square of the 4-momentum
transfer) have been obtained. Unfortunately, the statistical
errors are not in general appreciably smaller than in the earlier work.
This is presumably due to the  inherent noisiness associated
with lattice propagators of particles with non-zero three-momentum.
Just as for static-light 
mesons \cite{LEPAGE_LAT91,BLS_LAT91,EICHTEN_LAT91,LEPAGE_TASI},
the signal-to-noise ratio for mesons with non-zero momentum
falls exponentially because the average of the squared propagator
has non-zero overlap with a state which has less than twice
the meson energy. In this case, the overlap is with a
state of two mesons at rest.

One interesting issue concerns the value of the form factor $A_2(0)$
(or the ratio $A_2(0)/A_1(0)$).  Because of the size of the
errors, all  the new and old results agree within
$2 \sigma$.  The new work has, however, has helped to
change the qualitative picture.  Initially the ELC group
\cite{ELC_SEMIOLD} found an $A_2(0)$ consistent with zero
($0.19\pm0.21$ or, later, $0.4\pm0.4$) and with the initial experiment 
($0.0\pm0.2\pm0.1$) \cite{E691}.  On the other hand, the
BES \cite{BES}  result was distinctly different from
zero ($A_2(0)/A_1(0) = 0.70\pm0.16{}^{+0.20}_{-0.15}$).
The recent lattice calculations are now giving an $A_2(0)$
bounded away from zero: $A_2(0)/A_1(0) = 0.7\pm 0.4$ \cite{ELC_SEMINEW};
$A_2(0) = 0.67\pm0.44$ or $0.72\pm0.50$ \cite{STELLA} (on two
sets of configurations).  
A non-zero $A_2(0)$ is at present also found by experiment:
$A_2(0)/A_1(0) = 0.82\pm0.23\pm 0.11$ \cite{E653};
$A_2(0) = 0.44\pm 0.09$ \cite{WITHERELL}.

Although ref.~\cite{ELC_SEMINEW}
takes a look at the extrapolation to the $B$ meson, the other
lattice studies to date on 
heavy-light semileptonic form
factors have focussed exclusively on $D$ mesons.
The
issue of normalization of fields and currents when $am_Q\sim 1$ is thus not
as crucial here as for $f_B$. Still, it is significant numerically. 

Consider the normalization of the lattice
vector current for large $am$.
For the local current, $V_\mu^{\rm local}(x) = \bar \psi(x) \gamma_\mu \psi(x)$,
the arguments of refs.~\cite{KROMAC,LEPMAC} for normalizing the
quark field would just give a
normalization factor 
\begin{equation}
\label{eq:zvlocal}
\tilde Z^{\rm local}_V= 1-6\tilde\kappa + \cO(g^2) =
2\tilde\kappa e^{\tilde M_1a} + \cO(g^2) \ ,
\end{equation}
where the $\cO(g^2)$ terms are expected to be small since the
tadpoles are summed by using $\tilde\kappa$ and $\tilde M_1$.

The normalization of the conserved current,
$$\displaylines{
V_\mu^{\rm cons}(x) = {1\over2}\big[\bar \psi(x) (\gamma_\mu -1)U_\mu(x) 
\psi(x+\hat\mu)\hfill\cr
\qquad + \bar \psi(x+\hat\mu) (\gamma_\mu +1)U^\dagger_\mu(x) 
\psi(x)\big] \ ,\hfill }
$$
is more subtle.  In the mean field approximation,
it is not however hard to see how the current must
be normalized in the limit $am\to \infty$.  
Consider the matrix element 
\begin{equation}
\label{eq:vme}
\sum_{\vec x}\langle 0 | \psi(\vec x, t) V_\mu(\vec 0, t') \bar \psi(\vec 0, 0)
| 0 \rangle \ ,
\end{equation}
and compare the case $V_\mu=V_\mu^{\rm cons}$ to $V_\mu=V_\mu^{\rm local}$.
Because the quark just propagates in the time direction for large $ma$,
the difference between $V_0^{\rm cons}$ to $V_0^{\rm local}$ will 
simply be that the former has an extra factor of $u_0$ (the
mean field value of the explicit link in its definition) but is missing
a factor of $e^{-\tilde M_1a}$ (since the point splitting allows it
to skip one hop in the time direction). We thus expect
\begin{equation}
\label{eq:zvconszero}
\tilde Z^{{\rm cons,\ }\mu=0}_V \approx {e^{-\tilde M_1a} \over u_0}\tilde
Z_V^{\rm local}
\simeq 2\kappa 
\end{equation}
Eq.~(\ref{eq:zvconszero}) is actually exact; the conserved current obeys
a Ward identity which can be used to show that $\sum_{\vec x}
2\kappa V_0^{\rm cons}(\vec x, t)$ just counts the total 
charge \cite{LABRENZ}.  

The spatial components  $V_j^{\rm cons}$ are however normalized
differently.  Comparing eq.~(\ref{eq:vme}) for conserved
and local currents, we see that the former again has an extra factor of $u_0$
but is not missing the factor $e^{-\tilde M_1a}$, since no hops in
the time direction are saved.  We thus expect
\begin{equation}
\label{eq:zvconsj}
\tilde Z^{{\rm cons,\ }\mu=j}_V \approx {1 \over u_0}\tilde
Z_V^{\rm local}
= 2\kappa e^{\tilde M_1a}  + \cO(g^2)\ .
\end{equation}
Note that although the $\mu=0$ and $\mu=j$ components of $V_\mu^{\rm cons}$
enter the same Ward identity, this does not guarantee that they
are normalized the same.    The identity involves the divergence
of  $V_\mu^{\rm cons}$, rather than  $V_\mu^{\rm cons}$
itself, and one would need to take into account the difference between
$e^{ma} - 1$ and $ma$ for the discrete time derivative.

In the literature, the quantity
\begin{equation}
\label{eq:zvnonpert}
{\langle \rho | V_j^{\rm cons}|0\rangle \over 
\langle \rho | V_j^{\rm local}|0\rangle}  = {\tilde
Z_V^{\rm local} \over
\tilde
Z^{{\rm cons,\ }\mu=j}_V} \ ,
\end{equation}
where $\rho$ is a generic vector state, is sometimes used as
a ``non-perturbative definition'' of $Z_V^{\rm local}$.
However, the above arguments show that $\tilde
Z^{{\rm cons,\ }\mu=j}_V $ is not fixed by the Ward identity
and that the ``non-perturbative $Z_V^{\rm local}$'' should in fact
be roughly independent of the quark mass, unlike the
correct $Z_V^{\rm local}$ of eq.~(\ref{eq:zvlocal}).  Indeed, this
rough mass independence is found in simulations 
\cite{HEMCGC,ELC_SEMINEW,GUPTA}.  The independence 
is not evidence for a failure of the 
Kronfeld-Mackenzie ideas, as is suggested 
in refs.~\cite{ELC_SEMINEW,GUPTA}.  On 
the contrary, it is evidence in favor of ref.~\cite{KROMAC}.

One can go further.  Table 7 of ref.~\cite{ELC_SEMINEW} gives
the values of the ``non-perturbative $Z_V^{\rm local}$''
 from various comparisons
with the conserved current.  Using eqs.~(\ref{eq:zvconszero}),(\ref{eq:zvconsj})
and a tadpole-improved perturbative
calculation of $Z_V^{\rm local}$ with coupling $g_V(1/a)$,
I am able to (roughly) reproduce their numbers, as shown in
Table~\ref{tab:zv}.  I am neglecting the non-zero 3-momentum
that appears in the last two matrix elements.

\begin{table}
\caption{Values of ``$Z_V^{\rm local}$'' from Table 7 of ref.~[43],
and the values as calculated perturbatively using the
ideas of refs.~[31,14].
$J$ is the heavy-light
and $V$ is the light-light vector current, respectively.}
\label{tab:zv}
\centering
\begin{tabular}{|c|c|c|} \hline
matrix elem. & ref.~\cite{ELC_SEMINEW} &pert.th.  \\
\hline
$\langle V_1V_1\rangle$ & 0.659(3) & 0.70 \\
\hline
$\langle KJ_0D\rangle$ & 0.87(-) & 0.83 \\
\hline
$\langle KJ_1D\rangle$ & 0.75(1) & 0.70 \\
\hline
$\langle K_3^*J_1D\rangle$ & 0.66(2) & 0.70 \\
\hline
\end{tabular}
\end{table}

Some groups (\eg refs.~\cite{GUPTA,HEMCGC}) have
compared with experiment their results for
the vector meson decay
constant, $1/f_V$,  as a function of $M_P^2/M_V^2$ (the square of the
pseudoscalar to vector meson mass ratio).
Such comparisons can be very misleading since it is known
that Wilson quarks get much too small a value for the $M_V-M_P$
splitting for heavy quarks. Thus this is not a good way
to test various large-$am$ normalizations of the vector current.

Finally, 
the normalization factor $\sqrt{e^{M_1a}}$, which is not included
in the analysis of ref.~\cite{ELC_SEMINEW},
is as
as large as $1.3$ for their heaviest quark masses.  Thus the
extrapolation of form factors to the $B$ mesons in that paper
should be taken as qualitative only.

\section{FORM FACTORS FOR \bkg}

The penguin decay $b\to s \gamma$ \cite{MISIAK} provides an excellent 
test of the Standard Model (SM).  It is a short-distance decay
at the quark level (the loop with a virtual $t$-quark dominates),
so the inclusive rate can be calculated perturbatively.
In the SM, the branching ratio $BR(b\to s \gamma)$
is a slowly varying function of the top quark mass and is
essentially independent of the CKM angles in the 3-generation case.
These features make it very sensitive to physics beyond the SM.

The branching ratio for the exclusive decay
\bkg\ has been measured experimentally
to be $(4.5\pm1.5\pm0.9)\times 10^{-5}$ \cite{CLEO}.
For theory to make contact with this result, 
the ratio
\begin{equation}
R_{K^*}\equiv {\Gamma(B\to K^* \gamma) \over  \Gamma(b\to s \gamma)} \ ,
\end{equation}
is required.
Phenomenological evaluations of $R_{K^*}$ vary from $\sim 1\%$ to
$\sim97\%$, so the lattice can help out greatly here.

 In leading
logarithmic order one only one needs
to compute the form factors $T_i(q^2)$, defined by
$$\displaylines{
\langle K^*(k)| \bar s \sigma^{\mu\nu}q_\nu {(1+\gamma_5)\over 2} b|B(p) \rangle
=\hfill\cr 
\hfill\sum_{i=1}^3 c_i^\mu(p,k) T_i(q^2) \ ,\qquad}
$$
where the operator on the LHS arises from the short distance 
expansion of the penguin loop,
$q\equiv p-k$ is the photon momentum,
and the definitions of the coefficients $c_i^\mu$ can be found
for example in refs.~\cite{BHS,UKQCD_BKG}.  The form factors
obey $T_1(0)=T_2(0)$, and $T_3$ does not contribute
for a real photon, 
so $R_{K^*}$ is determined by $T_1(0)$.

After a first look on the lattice \cite{BHS_OLD},
two evaluations of the form factors have been attempted
recently \cite{BHS,UKQCD_BKG}.  An extrapolation
to the $B$ mass is required as usual.  The two groups
differ in how they perform the extrapolations. BHS \cite{BHS}
check that pole dominance relates $T_1(0)=T_2(0)$ to $T_2(q^2_{\rm max})$,
then extrapolate  $T_2(q^2_{\rm max})$ to the $B$ using
the relation derived from heavy quark effective theory:
\begin{equation}
\sqrt{M_P}T_2(q^2_{\rm max}) \sim A + {B \over M_P}\ ,
\end{equation}
and finally use pole dominance to get $T_2(0)$ from $T_2(q^2_{\rm max})$
at the $B$.  UKQCD \cite{UKQCD_BKG} calculate $T_1(0)$ directly
on the lattice (with some mild interpolation) and then extrapolate
$T_1(0)$ linearly in $1/M_P$ to the $B$.  

Each method has some disadvantages:  BHS use pole dominance over
a wider range ($q^2_{\rm max}/M_B^2 \approx 0.65$) than can be checked
with the current lattice data ($q^2_{\rm max}/M_P^2 \ltsim 0.3$).
UKQCD do a ``blind'' (\ie without theoretical
guidance) extrapolation of $T_1(0)$ to the $B$ mass.  Obviously
I prefer the BHS approach, but both methods are plausible.
The difference in the final results
gives some measure of the systematics:
BHS get $T_1(0)= 0.10 \pm 0.01\pm 0.03$;
UKQCD get  $T_1(0)= 0.15{}^{+.05}_{-.04}$ (assuming approximate
independence of the spectator quark mass).

Using their result for $T_1(0)$, BHS find
$R_{K^*}= (6.0\pm1.2\pm3.4)\%$.  Combined with the experimental
$BR$ for \bkg, this produces a $1\sigma$ bound which is, unfortunately,
uninteresting:
$m_t \gtwid 100$ GeV.  However, because $BR(b\to s\gamma)$
is a slowly varying function of $m_t$, it would not take much improvement
in the lattice and/or experimental results to produce a rather
stringent bound.

\section{RECENT DEVELOPMENTS ON $B_K$}

In the past year, there has been considerable progress in the
computation of $B_K$, the $K^0$-$\bar K^0$ mixing parameter.
First of all, the calculations of the $\cO(g^2)$ perturbative 
corrections to all
relevant lattice operators have been  
completed \cite{SHEARD,SHARPE_PERTOLD,PS,ISHIZUKA_PERT,SP}. These include
4-quark operators of the following types:
gauge-noninvariant (Landau gauge)  ``unsmeared'' 
($2^4$) \cite{SHARPE_PERTOLD,ISHIZUKA_PERT,SP} 
and ``smeared'' ($4^4$) \cite{PS,SP}, and 
gauge-invariant  ($2^4$ with links) \cite{SHEARD,ISHIZUKA_PERT}.
Without the perturbative corrections, results for $B_K$ from different
operators differ by as much as $\sim15\%$ ($\sim 7\sigma$) at
$\beta=6.0$ \cite{ISHIZUKA_LAT92,ISHIZUKA_PRL} and as
much as $\sim7\%$ ($\sim 5\sigma$) at
$\beta=6.2$ \cite{SHARPE}.
With the perturbative corrections included, results with
different operators agree within 
errors \cite{ISHIZUKA_LAT92,ISHIZUKA_PRL,SHARPE}.
An example is shown in Fig.~\ref{fig:kek}, taken from ref.~\cite{ISHIZUKA_PRL}.
Note 
that with the bare lattice 
coupling the disagreement between different operators is reduced
but not removed \cite{ISHIZUKA_LAT92};
full agreement occurs when one uses an
improved (``boosted'') coupling {\it \`a la} 
Lepage and Mackenzie \cite{LEPMAC}. (Tadpole improvement is not used here
because various operators involve different numbers of links.)
The 
results \cite{GKS,SHARPE_LAT91,ISHIZUKA_PRL}
also agree between groups.

\begin{figure}[tb]
\vspace{2truein}
\caption{Comparison [60] of $B_K$ before (open symbols)
and after (solid symbols) renormalization correction. The
superscript ``inv'' denotes gauge-invariant operators; ``non-inv,''
unsmeared Landau-gauge operators.}
\label{fig:kek}
\vspace{-0.15truein}
\end{figure}

The second advance involves the comparison of quenched and full
QCD results for $B_K$.  At the level of the statistics ($\sim2\%$),
no difference is found \cite{ISHIZUKA_PRL,KILCUP}.
The two dynamical quark masses in the full QCD simulations are degenerate
and roughly equal to
the average quark mass in the kaon  ($m_q \approx m_s/2$).
One can easily estimate, using the
known chiral logarithms \cite{BIJNENS,SHARPE_CHPT},
how large a difference one would expect in the full theory between
the simulated case and the physical situation with $m_d\approx m_u \approx 0$.
The result is a difference of only 3\% to 4\% \cite{SHARPE}.
This of course needs to be checked in simulations.  
Unfortunately, the quenched
theory cannot be used to estimate the difference between a
$B_K$ with degenerate quarks and one with a light $m_d$.  Quenched
chiral perturbation theory shows \cite{SHARPE_CHPT} that the
chiral logs for a $B_K$ with non-degenerate quarks (in contrast
to the degenerate case)
are different in the quenched and full theories.

The third advance is the understanding of the lattice
spacing errors.  At {\it Lattice '91}, the $B_K$ data
\cite{SHARPE_LAT91} showed
a strong dependence on $a$, and it was unclear how to
extrapolate to the continuum:  A linear extrapolation
in $a$ gave $B_K\alpha_s^{-2/9}\equiv \hat B_K=0.66(6)$; 
a quadratic, $\hat B_K=0.78(3)$.  This difference dominated
the systematic errors.  Sharpe \cite{SHARPE} has now shown 
that the lattice spacing errors are $\cO(a^2)$.  The basic idea
is simple:  
the staggered fermion action has corrections only at $\cO(a^2)$.
So corrections of $\cO(a)$ to $B_K$ could come only from the
dimension-6
weak operators themselves, through mixing with  
dimension-7 operators. However, a detailed enumeration shows that
no such operators exist with all the right symmetries:
hypercubic group, flavor symmetry, and individual axial
rotations of each flavor.

Once one knows the lattice spacing errors are  $\cO(a^2)$, it is straightforward
to include the perturbative corrections
and extrapolate existing data of 
Gupta, Kilcup and Sharpe \cite{SHARPE_LAT91} to the
continuum.  The result is \cite{SHARPE}
\begin{eqnarray}
\label{eq:bk}
&B_K({\rm NDR,2GeV}) =0.616\pm0.020\pm0.017 \\
&\hat B_K =0.825\pm0.027\pm0.023 \ ,
\end{eqnarray}
where the continuum $\alpha_s$ with $\Lambda^{(4)}_{\overline{\rm MS}}=300$MeV
is used to obtain the second equation.
The only errors not included above are those due to
quenching and to the use of degenerate quarks.  As discussed
above, the quenched and full theories  agree for degenerate
quarks at a scale roughly that of the quenched
$\beta=6.0$ ($a^{-1} \sim 2$GeV).  One needs
however to check that the theories continue to agree at
smaller lattice spacings. In addition, the effects of non-degenerate
quarks in the full theory need to be studied.  However, given the
size of the chiral logs, I would be surprised if this amounted
to more than a 5\% or 6\% error.  Thus if one at most doubles the systematic
errors 
in eqs.~(\ref{eq:bk}),(23), one has a result which is likely to be very
reliable.  This is a major accomplishment of lattice QCD.

I'd like to briefly mention one other result which has
implications for light-light weak matrix elements.
Kuramashi and collaborators \cite{KURAMASHI} have obtained results for 
$\pi$-$\pi$ scattering in QCD.  Their method makes possible the first
calculation of the $I=0$ scattering length.  The same method may ultimately
allow the computation of the $\Delta I=1/2$ $K\to\pi\pi$
weak amplitude, a computation which has bedeviled some of us in
lattice gauge theory
over the past ten years. 

\section{CONCLUSION}

The basic results have already been summarized in the introduction.
I would just like to add the observation that we owe a great
debt to Lepage and Mackenzie for showing how to
understand and control lattice perturbation theory \cite{LEPMAC}.
This has made possible many of the advances discussed above.

\section{ACKNOWLEDGEMENTS}

I thank 
 C.\ Allton,
 S.\ Collins,
 C.\ Davies,
 T.\ DeGrand,
 T.\ Draper,
 R.\ Gupta,
 S.\ G\"usken,
 E.\ Eichten,
 B.\ Hill,
A.\ Kronfeld,
J.\  Labrenz,
D.\ Leinweber,
 P.\ Mackenzie,
 G.\ Martinelli,
M.\ Okawa,
 A.\ Pich,
 D.\ Richards,
 C.\ Sachrajda,
H.\ Shanahan,
S.\ Sharpe,
 A.\ Soni,
 R.\ Sommer,
N.\ Stella,
 and
 A.\ Ukawa
for useful discussions. This work was supported in part
by the US Department of Energy under grants   DE2FG02-91ER40628
and DEFG0393-ER25186.


\begin{thebibliography}{9}
\bibitem{KENWAY}
I do not discuss the Isgur-Wise function; see instead 
R.\ Kenway, these proceedings.
\end{thebibliography}
\end{document}